\documentstyle[12pt]{article}
\begin{document}

\begin{flushright}
	                                         TUIMP-TH-95/63
\end{flushright}

\begin{center}
{\large\bf DETECTING THE INTERMEDIATE-MASS HIGGS \\
BOSON THROUGH THE ASSOCIATE \\
PRODUCTION CHANNEL $pp\rightarrow t\overline{t}HX\;\;$
\footnote{Work supported by the National Natural Science Foundation of
China and the Fundamental Research Foundation of Tsinghua University}}\\

\vspace{1.5cm}
  Hong-Yi Zhou \hspace{0.4cm} and \hspace{0.4cm} Yu-Ping Kuang\\
       CCAST (World Laboratory), \hspace{0.2cm}
      P. O.\hspace{0.2cm}  Box 8730, Beijing 100080, China,\\
      and Institute of Modern Physics and Department of Physics,\\Tsinghua
      University, Beijing 100084, China \footnote{Mailing address} \\

\vspace{2cm}

		           {\large \bf ABSTRACT}\\
\end{center}
	We examine the detection of the intermediate-mass Higgs boson (IMH)
   	at LHC through the associate production channel $pp\rightarrow
	t{\bar t}HX\rightarrow l\gamma\gamma  X^{\prime}$. It is shown that
	by applying kinematic cuts or b-tagging on the final state jets,
	the main backgrounds of $W(\rightarrow l\nu)+\gamma+\gamma+(n-jet)$
     can be reduced substantially without significant loss of signals.
     It is possible to detect the IMH at LHC through the $pp\rightarrow
	t{\bar t}HX$ channel using a modest photon detector with mass
	resolution $\sim 3\%$ of the photon pair invariant mass.

\newpage
{\flushleft \bf I. Introduction}

The most mysterious part of the Standard Model (SM) is the symmetry breaking
sector. The search for the Higgs boson which is responsible for the symmetry
breaking has become one of the main tasks  in high energy physics. LEP II
can search for a Higgs boson of mass up to $80~GeV$ \cite{SLWU}. For $m_H
\geq 80~GeV$, the detection of the Higgs boson will be left to the CERN
Large Hadron Collider (LHC), Next Linear Collider (NLC) or $ep$, $e\gamma$
colliders. A Higgs boson in the intermediate-mass range, $80~GeV\leq m_H\leq
140~GeV$, is shown to be particularly difficult for LHC to detect. Recent
studies show that a SM Higgs in this region can be detected at LHC\cite
{Stange}, LEP$\bigotimes$LHC $ep$\cite{Leil} and $TeV$ $e\gamma$
colliders\cite{Cheung} via WH/ZH production with leptonic or hadronic decays
of W/Z and $H\rightarrow b{\bar b}$ or $H\rightarrow\gamma\gamma$\cite
{Stirling}. There are also proposals of detecting a SM intermediate-mass
Higgs boson (IMH)  through  $t{\bar t}H$ production with inclusive final
state signals of $l\gamma\gamma$
\cite{Marciano}\cite{Gunion}\cite{Kunszt}\cite{Maina}\cite{Summers}. However,
our recent study\cite{ZK} shows that there exist difficulties due to the
large reducible background processes $pp\rightarrow W(\rightarrow l\nu)+
\gamma+\gamma+(n-jet)+X~(n=0,1,2,3,4)$ with which the inclusive
$l\gamma\gamma$ detection of the IMH in the $t{\bar t}H$ production needs a
high level photon detector with photon pair invariant mass
($M_{\gamma\gamma}$) resolution of $\sim 1\%$. In Ref.\cite{ZK}, various
tree-level contributions have been taken into account. In view  of the fact
that the infrared divergences (IFD) in the tree-level collinear- or soft-
gluon emission diagrams are cancelled by the IFD in loop diagrams, one may
worry about the uncertainty in Ref.\cite{ZK} due to ignoring the probable
residual cancellation effect if the jet-$p_T$ cut is not large enough
\cite{Summers}. However, taking the largest one-jet process as an example,
our result shows that, with the cut $p_T>30~GeV$, the main contribution
comes from the gluon-quark fusion part which does not contain IFD, while the
gluon emission contribution is only about one-tenth of it. Hence such kind
of probable uncertainty does not really affect the main feature of the
results in Ref. \cite{ZK}. Therefore the backgrounds $W(\rightarrow l\nu)+
\gamma+\gamma+(n-jet)+X$ with $n\geq 1$ should really be taken seriously.
Unfortunately, most of the recent papers concerning the $t{\bar t}H$
production channel, including the most recent realistic Monte-Carlo
simulations by the CMS and ATLAS collaborations\cite{CMSATLAS}, have not
carefully taken such backgrounds in to account. Even if an IMH can be
detected at LHC via WH/ZH production, it is still worthy to observe
the $t{\bar t}H$ production to explore the coupling of the Higgs boson to
the top quark. It is then our purpose in this paper to investigate the
possibility of reducing these backgrounds and obtaining a large signal to
background  ratio ($S/B$) at LHC with a modest photon detector of $\sim 3\%$
$M_{\gamma\gamma}$ resolution.

 In Sec II we analyze the backgrouds to the SM $t{\bar t}H$ production with
inclusive $l\gamma\gamma $ final states at LHC and propose the methods of
reducing them. We then present the number of events of signals and
backgrounds without and with b-tagging of 100\% efficiency. In Sec III we
give our discussions and conclusions .

{\flushleft \bf II. Background Reduction and Results}

	If the Higgs boson is detected via inclusive final states
$l\gamma\gamma $ from $pp\rightarrow t{\bar t}HX$ production, the final
states will contain $0-4$ jets from $t{\bar t}$ decays, e.g. $t{\bar t}
\rightarrow WbW{\bar b}\rightarrow l\nu jjb{\bar b}$. Therefore the processes
$pp\rightarrow W(\rightarrow l\nu)+\gamma+\gamma+(n-jet)+X~(n=0,1,2,3,4)$
will be the reducible backgrounds to the Higgs boson signal. After applying
isolation cuts it is found that apart from the above mentioned backgrounds
the main remained background is from the irreducible $t{\bar t}\gamma\gamma$
process which is not significant\cite{Kunszt}\cite{Summers}. Although the
$n\geq 1$ contributions seem to be of higher order of QCD with respect to the
$n=0$ $W\gamma\gamma$ process, our explicit calculations\cite{ZK} show that
they are surprisingly larger than the $W\gamma\gamma$ background, partly due
to the appearance of channels of $qg$ and $gg$ in the initial states. Our
results are also consistent with the result $\sigma(W+(3-jet))>
\sigma(t{\bar t})$\cite{Cavanna} which implies $\sigma(W+\gamma+\gamma+
(3-jet))>\sigma(t{\bar t}\gamma\gamma)$. Therefore, our main task is to
reduce the backgrounds from the $pp\rightarrow W(\rightarrow l\nu)+\gamma+
\gamma+(n-jet)+X$ processes. Inspired by the reduction of $W+(n-jet)$
backgrounds to the $t{\bar t}$ signal\cite{Reya}, we investigate the
possibilities of reducing the $W+\gamma+\gamma+(n-jet)$ backgrounds to
$t{\bar t}H$ signal in the  $l\gamma\gamma$ mode.

There are some notable features of the $t{\bar t}H$ signal events. First,
almost 100\% events contain at least two jets and 80\% contain more than two
jets due to the heaviness of the top quark \cite{Summers}. Secondly, the
final state signal jets contain contribution from $t\rightarrow Wb\rightarrow
jjb$ decay which may allow us to reconstruct W and t from the detected jets.
The third feature is that there are b quark jets in the signal from $t{\bar t}
\rightarrow WbW{\bar b}$ decay which can be tagged.

The first thing we do is to require at least three jets in the final states.
As we have mentioned, there are still $80\%$ signals satisfying this
requirement, while the largest $n=1,2$ backgrounds are eliminated, i.e. the
remaining backgrounds are only $t{\bar t}\gamma\gamma$ and the $n=3,4$ ones.
Unfortunately we can not explicitly calculate the $n=3,4$ backgrounds due to
the large number of Feynmann diagrams. For example, the number of diagrams is
1758 with the external lines $Wqq^{\prime}ggg\gamma\gamma$. Our previous
estimate shows that they are important\cite{ZK}. However, we can make an
approximate estimate of the $n=3,4$ backgrounds from the calculated
$pp\rightarrow W+(n-jet)$ cross-sections\cite{Reya}. Our calculations show
that $\sigma(W+\gamma+\gamma+(2-jet))/\sigma (W+\gamma+\gamma+(1-jet))$ is
about $0.7$ at LHC. This number is close to the ratio $\sigma(W+(2-jet))/
\sigma(W+(1-jet))=80/52\approx 0.7$ given in Ref.\cite{Reya}. This implies
that the emission of two extra photons does not affect the ratio much. So it
is likely that cases containing more jets may have the similar situation.
Then we can expect that $\frac{\sigma(W+\gamma+\gamma+(3-jet))}{\sigma(W+
\gamma+\gamma+(2-jet))}$  and $\frac{\sigma(W+\gamma+\gamma+(4-jet))}
{\sigma(W+\gamma+\gamma+(2-jet))}$ are close to $\frac{\sigma(W+(3-jet))}
{\sigma(W+(2-jet))}$ and $\frac{\sigma(W+(4-jet))}{\sigma(W+(2-jet))}$ which
are $24/52\approx 1/2$ and $8.6/52\approx 1/6$, respectively, according to
Ref.\cite{Reya}.

What we are going to do next is to impose certain kinematical cuts on the
jets to further enhance the signal to backgrounds ratio. As there is large
probability that two jets in the signal come from W decays, we impose a cut
on the two-jet invariant mass $m_{jj}= m_W\pm\delta m$ with a resolution
$\delta m$. This will further reduce the $n=3,4$ backgrounds relative
to the signal. We define {\it cut efficiency} as the ratio of the cross-
section with the cut to that without the cut. Let $\epsilon_2$ be the cut
efficiency of the two-jet invariant mass cut in the $n=2$ process which will
be calculated in the way given in Ref.\cite{ZK}. In the $n=3$ process, there
are 3 combinations of two-jet pairs. A simple estimate regarding them as
independent events leads to a cut efficiency $\epsilon_3=3\epsilon_2$. In
the $n=4$ case, the simple estimate gives the cut efficiency $\epsilon_4=
6\epsilon_2$. Therefore, with the above estimated $n=3,4$ background cross-
sections, the cross-section of $n=3,4$ after this cut will be about the same
as that of $n=2$. If we can measure the top quark mass more accurately in the
future experiments at Tevatron or LHC, we can further require a third jet
combined with the two satisfying $m_{W}-\delta m<m_{jj}<m_{W}+\delta m$ to
form $m_{jj{j}^{\prime}}$ and $m_{t}-\delta m<m_{jj{j}^{\prime}}<m_{t}+\delta
m$, reflecting that the three jets come from the $t$ decays which has a
large probability in the signal but not in the $n=3,4$ backgrounds.
In this case, we get the combined cut efficiencies ${\epsilon}^{\prime}_3=
3\epsilon^2_2$, ${\epsilon}^{\prime}_4=12\epsilon^2_2$. In obtaining this
result, we have simplely assumed that the cut efficiency of one combination
satisfying $m_{jj{j}^{\prime}}=m_{t}\pm\delta m$ cut is $\epsilon_2$. We
shall discuss this in the next section.  Note that we count the event only
once if there is one combination satisfying the cuts regardless of the number
of combinations.

In our calculations , we use the following parameters and parton distribution:
\begin{equation}\begin{array}{l}
	\sqrt{s}=14~TeV,\;\;\;\int{\cal L}dt=100fb^{-1},\;\;\;
        M_{\gamma\gamma }\;\;resolution=3\%,\;\;\;m_t=176~GeV;\\
	for\;\; q\overline{q},\;gg\rightarrow t\overline{t}H\;\; and\;\;
	q\overline{q},\;gg\rightarrow t\overline{t}\gamma\gamma :\;\;\;
        Q^2=\hat{s};\\
	for\;\;W\gamma\gamma +2-jet:\;\;\;Q^2=m_W^2;\\
	MRS\;\;Set \;A^\prime \cite{St},\;\;\;\Lambda=231\;MeV \;.
\end{array}\end{equation}
     As in Ref\cite{Summers},
the following cuts are used for the final state particles:
\begin{equation}\begin{array}{c}
    p_T(l,\gamma)>20\;\;GeV \;\;|\eta(l,\gamma,jet)|<2.5 ,\\
    \Delta R(jet_1,jet_2)>0.4,\;\;\Delta R(\gamma_1,\gamma_2)>0.4 \;\;,\\
    \Delta R(l,\gamma)>0.4,\;\; \Delta R(\gamma,jet)>0.4\;\;,\\
    \Delta R(l,jet)>0.4,\;\; 0<M_{\gamma\gamma }<200\;\; GeV,
\end{array}\label{llbb}\end{equation}
where $\Delta R\equiv\sqrt{\Delta\phi^2+\Delta\eta^2}$. We allow the
transverse momenta $p_T$ of jets  to vary as given in the tables.

	The results of the $m_{jj}=m_W\pm\delta m$ and
$m_{jj{j}^{\prime}}=m_t\pm\delta m$ cuts are presented in TABLE I and TABLE
II corresponding to $\delta m=10~GeV$  and $\delta m=20~GeV$, respectively.
Our results show that the $S/B$ ratios are improved. A good $m_{jj}$
resolution of $\delta m=10~GeV$ will give a clear signal even if we use only
the $m_{jj}=m_W\pm\delta m$ cut. The $S/B$ ratios are not so good when
$\delta m=20~GeV$ if only the $m_{jj}=m_W\pm\delta m$ cut is applied. Note
that the $S/\sqrt B$ values with the combined cuts in TABLE I and TABLE II
are of the same level as those in Ref.\cite{CMSATLAS} wherein the $n\geq 1$
backgrounds are not taken into account.

There is another method of reducing the $pp\rightarrow W(\rightarrow l\nu)
+\gamma+\gamma+(n-jet)+X~(n=1,2,3,4)$ backgrounds. It is the use of b-
tagging in the final jets requiring at least one b-jet in the final jets
which will lead to significant reduction of the reducible backgrounds as
in the case considered in Ref.\cite{Reya}. The main backgrounds are then from
the irreducible $t{\bar t}\gamma\gamma$ process and the reducible $W+\gamma+
\gamma+(1,2,3,4)-jet$ processes with jet(s) faking the b-jet(s). We use the
approximations of $\sigma (W+\gamma+\gamma+(3-jet))\sim\sigma(W+\gamma+
\gamma+(2-jet))/2$, $\sigma(W+\gamma+\gamma+(4-jet))\sim\sigma(W+\gamma+
\gamma+(2-jet))/6$  and a level of 1\% $jet\rightarrow b$ to estimate this
latter background. There are also possible backgrounds coming from $W\gamma
\gamma b{\bar b}$ and $W\gamma\gamma c{\bar b}$. The former is a subprocess
of $W+\gamma+\gamma+(2-jet)$ in diagrams with four external quark lines
like $q_1{\bar q}_2\rightarrow W\gamma\gamma q_3{\bar q}_3$, $q_1{\bar q}_3
\rightarrow W\gamma\gamma q_2{\bar q}_3$. According to our calculation,
this four quark processes contribute only 1/10 of the total $W+\gamma+\gamma+
(2-jet)$, and $W+\gamma+\gamma+b{\bar b}$ contributes at most about 1/10 to
the four quark processes. Therefore, this background will not exceed that
of the processes with jet faking b. $W\gamma\gamma c{\bar b}$ process
is a subprocess of $gg\rightarrow W\gamma\gamma q_1{\bar q}_2$ which is also
about 1/10 of $W+\gamma+\gamma+(2-jet)$ and a subprocess of the four quark
processes. For $q_1=c,q_2=b$, there are additional CKM or heavy flavor parton
distribution suppressions. These make this background negligible. Although
the efficiency $\epsilon_b$ of b-tagging at present is only $\sim 0.4$, there
may be possibility of improvement. We present the result in TABLE III with an
extreme case of $\epsilon_b=1$ for reference.

{\flushleft \bf III. Discussions and Conclusions}

Our results of applying $m_{jj}$ and $m_{jj{j}^{\prime}}$ cuts
are obtained with the simple estimate of the relation between the cross-
sections and the cut efficiencies of $n=3,4$ to those of $n=2$. Therefore
there are uncertainties. A factor of two uncertainty of the $n=3,4 $
backgrounds will not cause any problem if we apply both $m_{jj}$ and
$m_{jj{j}^{\prime}}$ cuts. Actually, the above estimate gives already an
over estimate of the possible backgrounds due to  the following fact.
As a check, we have calculated the cut efficiency for $m_{jj}\sim m_W$ plus
$m_{jj{j}^{\prime}}\sim m_t$ cuts of the $W+(3-jet)$ process by using the
program PAPAGENO. The result shows that the cut efficiency is actually much
smaller than $3\epsilon_2^2$.

Also the above estimate does not include any detection efficiencies of the
jets. But this will have no influence on the $S/B$ ratios since both the
signal and background are affected in the same way. In the b-tagging case,
we see from TABLE III that a realistic $\epsilon_b\sim 0.4$ still gives 6-8
signal events. These events might be too low for detection if some further
detection efficiencies are included. But it can be overcome by increasing
the integrated luminosity, say, to about $150fb^{-1}$.

In conclusion, an IMH can be detected at LHC in the mode $l+\gamma+\gamma+
(n-jet)$ from the $t{\bar t}H$ production with a modest photon detector of
photon invariant mass resolution $3\%$ when we use both the $m_{jj}\sim m_W$
and the $m_{jj{j}^{\prime}}\sim m_t$ cuts or b-tagging on the final state
jets if the b-tagging efficiency can be improved. When the jet mass
resolution can reach within $10~GeV$ ($\delta m =10~GeV$), we can detect an
IMH by using only $m_{jj}\sim m_W$ cut.

\vspace{0.8cm}
 We are grateful to I.Hinchliffe for provide us the program PAPAGENO which
we used in our calculations.

\newpage
\parindent=0pt
{\bf TABLE I.} Signal and background events after applying $m_{jj}=m_W\pm
\delta m$ and $m_{jj{j}^{\prime}}=m_t\pm\delta m$ cuts. Number of jets ($n_j$)
$\geq 3$, $p_T(jet)\geq 30~GeV$, $\delta m= 10~GeV$. The cut efficiency is
$\epsilon_2=0.053$ .

\begin{center}
\begin{tabular}{|c|c|c|c|c|c|c|}
\hline
cuts & $m_H~(GeV)$ & $t\overline{t}H$ & $t\overline{t}\gamma\gamma $
& \parbox[t]{3.0cm}{$
W\gamma\gamma +3-jet$\\ + $W\gamma\gamma +4-jet$} & \parbox[t]{2.4cm}
{Total\\ backgrounds} & $S/\sqrt{B}$  \\
\hline
$m_{jj}=m_W\pm\delta m$ & 70 & 7.1 & 0.7 & 2.0 & 2.7 & 4.3\\
 & 100 & 9.5 & 0.8 & 2.8 & 3.6 & 5.0\\
 & 130 & 6.9 & 0.7 & 1.4 & 2.1  & 4.8\\
\hline
$m_{jj}=m_W\pm\delta m$ & 70 & 5.5 & 0.5 & 0.1 & 0.6 & 7.1\\
plus  & 100 & 7.3 & 0.6 & 0.2 & 0.8 & 8.2\\
$m_{jj{j}^{\prime}}=m_t\pm\delta m$ & 130 & 5.3 & 0.6 & 0.2 & 0.8 & 5.9 \\
\hline
\end{tabular}
\end{center}

\newpage

{\bf TABLE II.}  Signal and background events after applying $m_{jj}=m_W\pm
\delta m$ and $m_{jj{j}^{\prime}}=m_t\pm\delta m$ cuts. $n_j\geq 3$,
$p_T(jet)\geq 30~GeV$, $\delta m= 20~GeV$. The cut efficiency is
$\epsilon_2=0.12$ .

\begin{center}
\begin{tabular}{|c|c|c|c|c|c|c|}
\hline
cuts & $m_H~(GeV)$ & $t\overline{t}H$ & $t\overline{t}\gamma\gamma $
& \parbox[t]{3.0cm}{$
W\gamma\gamma +3-jet$\\ + $W\gamma\gamma +4-jet$} &
 \parbox[t]{2.4cm}{Total\\ backgrounds} & $S/\sqrt{B}$\\
\hline
$m_{jj}=m_W\pm\delta m$ & 70 & 7.8 & 0.7 & 4.3 & 5.0 & 3.5\\
 & 100 & 10.5 & 0.9 & 6.3 & 7.2 & 4.0\\
 & 130 & 7.6 & 0.8 & 2.9 & 3.7 & 4.0 \\
\hline
$m_{jj}=m_W\pm\delta m$ & 70 & 5.8 & 0.6 & 0.8 & 1.4 & 5.0\\
plus  & 100 & 7.6 & 0.7 & 1.0 & 1.7 & 6.0\\
$m_{jj{j}^{\prime}}=m_t\pm\delta m$ & 130 & 5.6 & 0.7 & 1.2 & 1.9 & 4.1 \\
\hline
\end{tabular}
\end{center}

{\bf TABLE III.}  Signal and background events after requiring
at least one b-jet in the final states in addition to $l\gamma\gamma$.
$p_T(jet)\geq 20~GeV$.  $W\gamma\gamma+(1,2,3,4)-jet$ with jet faking
b-jet events are estimated with a level of 1\% $jet\rightarrow b$.

\begin{center}
\begin{tabular}{|c|c|c|c|c|}
\hline
$m_H~(GeV)$ & $t\overline{t}H$  signal & $t\overline{t}\gamma\gamma$
& $W\gamma\gamma+(1,2,3,4) jets$
& Total backgrounds\\
70 & 16.2 & 1.6 & 1.1 & 2.7\\
100 & 21.6 & 1.8 & 1.5 & 3.3\\
130 & 16.0 & 1.7 & 1.5 & 3.2\\
\hline
\end{tabular}
\end{center}

\end{document}